\documentclass[12pt,preprint]{aastex}

\usepackage{emulateapj5,psfig}
\usepackage{psfig}

\newenvironment{inlinefigure}{
\def\@captype{figure}
\noindent\begin{minipage}{0.999\linewidth}\begin{center}\small}
{\end{center}\end{minipage}\smallskip}

\def\gsim{~\rlap{$>$}{\lower 1.0ex\hbox{$\sim$}}}
\def\lsim{~\rlap{$<$}{\lower 1.0ex\hbox{$\sim$}}}

\def\kms{\,\hbox{km}\,\hbox{s}^{-1}}

\begin{document}

\title{Galaxies under the Cosmic Microscope: A GMOS study of the
  lensed disk-galaxy $\#$289 in A\,2218}

\shorttitle{Galaxies Under the Cosmic Microscope}

\author{
A.\,M.\ Swinbank,\altaffilmark{1}
J.\, Smith,\altaffilmark{2} 
R.\,G.\ Bower,\altaffilmark{1} 
A.\ Bunker,\altaffilmark{2} 
I.\ Smail,\altaffilmark{1} 
R.\, S.\ Ellis,\altaffilmark{3}
Graham\ P.\ Smith,\altaffilmark{1,3}
J.-P.\ Kneib,\altaffilmark{3,4} 
M.\ Sullivan,\altaffilmark{1}
J.\ Allington-Smith \altaffilmark{1}
}

\setcounter{footnote}{0}

\altaffiltext{1}{Department of Physics, University of Durham, South
  Road, Durham DH1 3LE, UK -- Email: a.m.swinbank@dur.ac.uk}
\altaffiltext{2}{Institute of Astronomy, University of Cambridge,
  Madingley Road, Cambridge, CB3 0HA, UK}
\altaffiltext{3}{California
  Institute of Technology, MC 105-24, Pasadena, CA 91125, USA}
\altaffiltext{4}{Observatoire Midi-Pyr\'en\'ees, 14 Avenue E.\,Belin,
  31400 Toulouse, France} 

\setcounter{footnote}{0}

\begin{abstract}
  
  In this letter, we exploit the gravitational potential of the rich
  cluster A\,2218 as a magnifying glass. We demonstrate that the
  magnification due to the cluster allows us to observe distant
  background galaxies at a comparable level of detail to galaxies at
  $z\sim0.1$.  Using the GMOS Integral Field Unit on Gemini North we
  observed the spatially-resolved [O{\sc ii}]\,$\lambda$3727 emission
  line spectrum for a lensed disk-galaxy at z=1.034.  Using a detailed
  model for the cluster mass distribution, we are able to correct for
  the lensing by the cluster and reconstruct the source morphology.  We
  find that the overall magnification is a factor of $4.92 \pm 0.15$,
  and the rest-frame absolute $I$-band magnitude is M$_{I}^{rest}$ =
  $-22.4 \pm 0.2$, where the error bars include conservative estimates
  of the uncertainty in the source-plane reconstruction.  The
  inclination-corrected circular velocity is $206 \pm 18\kms$.  The
  galaxy lies very close to the mean Tully-Fisher relation of
  present-day spirals.  Although our results are based on a single
  object, they demonstrate that gravitational lensing can be viably
  used to make detailed studies of the evolution of the structure of
  distant field galaxies.

\end{abstract}


\section{Introduction}
The deflection of light-rays by the deep gravitational potential of
galaxy clusters can be harnessed to greatly increase the effective
collecting area of astronomical telescopes. Such ``gravitational
telescopes'' magnify the images of background objects allowing us to
study faint distant galaxies in a level of detail that would simply not
be possible by conventional means (Smail et al.\ 1996; Franx et al.\ 
1997; Teplitz et al.\ 2000; Ellis et al.\ 2001; Campusano et al.\ 2001;
Smith et al.\ 2002).  In particular, many of the selection biases that
require us to only observe bright, extended galaxies are removed,
allowing us to test models for galaxy formation in a much fairer way
than otherwise possible.

In this paper, we will concentrate on the dynamics of galaxies seen at
$z\sim1$ by measuring the rotation speeds of gas in these systems
through the [O{\sc ii}]\,$\lambda$3727 emission line.  By observing the
distant ($z > 1$) universe, we sample cosmic history at a time before
many present-day stars had formed.  The ``classical'' galaxy formation
model, where the formation of a dark halo preceeds the formation of
the galaxy's disk (with the disk subsequently growing over an extended
period of time through the gradual accretion of gas from the halo, eg.\ 
Eggen et al.\ 1962; Larsen et al.\ 1980; Sandage\ 1990), predicts that
the stellar masses at a given circular velocity will be much lower at
$z=1$ (by about a factor of 2) than at the present-day.  In contrast,
CDM hierarchical models predict that galaxies would follow a similar
stellar mass versus circular velocity correlation at all redshifts.
Previous work in this area has, by necessity, concentrated on galaxies
which are intrinsically bright and mostly at $z<0.5$ (Vogt et al.\ 
1997; Verheijen 2001; B\"ohm et al.\ 2002; Ziegler et al.\ 2003), with
only a few galaxies being studied at higher redshift (Vogt et al.\ 
1996; Milvang-Jensen et al.\ 2002; Barden et al.\ 2003).  The results
suggest an increase in B-band luminosity for a given circular velocity
(Vogt et al.\ 2002; Kannappan et al.\ 2002) -- suggesting a preference
for hierarchical formation models.

Here we demonstrate the feasibility of using lensing to study the
evolution of the Tully-Fisher relation at much higher redshift.  We use
the GMOS IFU on Gemini-North to target the gravitationally lensed
$z=1.034$ arc (object \# 289 -- we use the same numbering system as
Pell\'o et al.\ (1992) in the cluster A\,2218.  We use a cosmology with
H$_{o} = 75\kms$Mpc$^{-1}$ and $q_{o} = 0.05$ to be consistent with
Vogt et al.\ (1997).  We note that for a cosmology with
$\Omega_M=0.3$, $\Omega_{\Lambda}=0.7$, the results are very similar
(the physical distances are only 6\% larger and the luminosities 12\%
brighter).

\section{Observations, Analysis $\&$ Results}

The galaxies lensed by the A\,2218 cluster were first studied by
Pell\'o et al.\ (1988), and later a number of arcs and arclets were
discovered in the cluster core (Kneib et al.\ 1996; Ebbels et al.\ 
1998; Ellis et al.  2001).  Spectroscopy of the arcs in the cluster
have confirmed a redshift for arc \# 289 of 1.034 (Pell\'o et al.\ 
1992; Ebbels et al.\ 1998).
 
%
%
\begin{figure*}[tbh]   
\centerline{\psfig{file=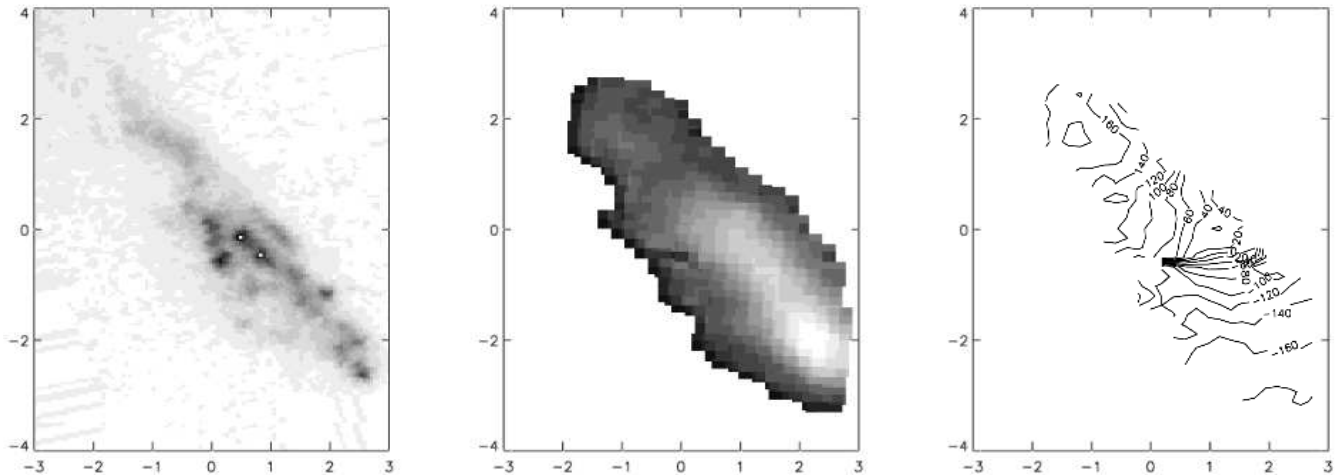,width=7in,angle=0}}   
\caption{Left: Arc \#289 in A\,2218 generated by combining 
  the {\it HST} WFPC2 $B_{450}$,$V_{606}$ and $I_{814}$ drizzled
  images.  (Middle) The [O{\sc ii}]\,$\lambda$3727 emission map of the
  arc measured from our IFU observations.  The distribution of [O{\sc
    ii}] emission agrees well with the UV flux seen in the left panel
  (the seeing for the observations was 0.7$''$).  (Right) The contour
  map of the derived velocity field of the galaxy.  The scale is marked
  in arcseconds, North is up and East is left.}
\label{fig:obs_images}
\end{figure*}   


\subsection {HST Observations and Lens Model}

The cluster was first observed by the Hubble Space Telescope {\it
  (HST)} WFPC2 in September 1994 in the R$_{702}$ filter for a total of
6.4\,ks (see Kneib et al.\ 1996).  It was subsequently observed again
in January 2000 using the B$_{450}$,V$_{606}$ and I$_{814}$ filters for
totals of 12.0, 10.0 and 12.0ks respectively (discussed in Smail et
al.\ 2001)\footnote {Based on observations made with the NASA/ESA {\it
    Hubble Space Telescope} which is operated by STSCI for the
  Association of Universities for Research in Astronomy, Inc., under
  NASA contract NAS5-26555.}.

Arc\#289 is a fairly blue disk-galaxy at $\alpha$ =
16$^h$35$^m$55$^s$.1, $\delta$ = +66$^o$11'51''.0 (J2000) lying in a
saddle between the central dominant galaxy of the cluster ($\#$ 301)
and a sub-clump of elliptical galaxies (\# 307 \& \# 292,
Figure~\ref{fig:obs_images} in Smail et al. 2001).  The {\it HST}
images of the galaxy show a large amount of internal structure with the
knotted, disk-like morphology resembling a late-type galaxy (especially
the $B$-band flux which samples the rest-frame UV and is therefore
dominated by the star-forming {H\sc ii} regions --
Figure~\ref{fig:obs_images} - left).  We calculate the magnitude of the
arc in various passbands in Table 1 by using the {\sc iraf imsurfit}
package with the sky estimated from a 2$^{\rm nd}$ order polynomial
surface fit to regions around the galaxy.  We then use {\sc sextractor}
(Bertin \& Arnaud 1996) to estimate the residual background within the
frame.

%
%
\begin{center}
\hspace*{-0.7cm}{\scriptsize
{\centerline{\sc Table 1: Photometric properties of \#289}}
\smallskip

\hspace*{-0.7cm}\begin{tabular}{lccc}
\hline\hline
\noalign{\smallskip}
Filter & Magnitude & {\it b/a} & inclination\\
\hline
\noalign{\smallskip}
$B_{450}$  & $21.66 \pm 0.04$ & $0.37$  & $ 68 $\\
$V_{606}$ & $20.86 \pm 0.10$  & $0.41$  & $ 66 $\\
$R_{702}$ & $20.53 \pm 0.04$  & $0.40$  & $ 66 $\\
$I_{814}$ & $19.79 \pm 0.05$ &$0.53$ & $ 58 $\\
$J$ & $18.63 \pm 0.06$ & $ - $ & $ - $\\
$K$ & $17.32 \pm 0.06$ &$ - $ & $ - $\\
$IFU $ & $ - $ & $ 0.5184 $ & $ 58 $\\
\hline
\end{tabular}
\vspace{0.2cm}

\hspace*{-0.0cm}\begin{tabular}{l} Aperture magnitudes of the arc are
 in the Vega-based system.\cr 
 Inclination is computed in the source (reconstructed) frame \cr
\end{tabular}
}
\end{center}
\vspace{0.1cm}


To correct for the distortion and magnification of the galaxy image by
the cluster lens we need to employ a lens model.  A detailed mass model
for A\,2218 was originally developed by Kneib et al.\ (1996) from {\it
  HST} imaging and this has been updated by Ellis et al.\ (2001) and
Smith et al.\ (2003).  In particular, Smith (2002, see also Smith et
al.\ 2003) incorporated all of the strong lensing constraints to
produce one of the best constrained strong lensing clusters, with three
spectroscopically confirmed multiple-image systems.  The precision of
this lens model makes this cluster and ideal gravitational telescope
with which to study (and reconstruct) the detailed properties of high
redshift galaxies.

While the giant blue arc \# 289 is highly distorted at the northern end
(Fig. 1a), extending across the halo of the cluster galaxy \# 244, the
majority of the source lies outside the caustic producing a highly
magnified, weakly sheared image.  The southern end of the arc does not
suffer the same strong shear, and is not so distorted although it has a
higher surface brightness than the northern end.  Exploiting the fact
that gravitational lensing conserves surface brightness, we compute the
magnification in each band by using the transformation between the
sky-plane and source-plane coordinates and compare the flux in the
(reconstructed) source and (observed) sky frames
(Fig.~\ref{fig:source_image} - left).  We use Smith's (2002) model of
A\,2218 to compute the magnification ($\mu$) of \# 289, obtaining a
mean luminosity weighted magnification of $\mu=4.92$ across the whole
galaxy image.  The spatial variation of surface brightnesses may vary
between filters due to differing spatial distributions of stellar
populations (and dust) in the lensed galaxy, thus causing the
magnification to vary by $\pm0.15$ depending on the band.  The
statistical error bar on $\mu$ is derived from the family of lens
models which adequately reproduce the multiply images arcs in the
cluster.  We estimate the 1--$\sigma$ uncertainty by perturbing the
parameters of the best fit lens model such that $\Delta\chi^{2}=1$.
For each model we recompute the magnification of \#289.  The error
corresponds to the largest variation in magnification that we found and
is converted into a conservative estimate of $\pm0.05$ magnitudes in
the source plane photometry of \#289.

\subsection{Ground-based Imaging}

Additional constraints on the spectral energy distribution (SED) of the
galaxy comes from the near-infrared imaging of this field using the
INGRID camera (Packham et al.\ 2003) on the William Herschel Telescope
\footnote {Based on observations made with the William Herschel
  Telescope operated in the island of La Palma by the Isaac Newton
  Group in the Spanish Observatorio del Roque de los Muchachos of the
  Instituto de Astrofisica de Canarias.}.  The reduction and analysis
of these data is described by Smail et al.\ (2001).  With our
multi-wavelength photometry (Table~1) we construct an SED of this very
blue galaxy whose colors indicate a current to past star formation rate
consistent with that seen for late-type spirals at the present day.
The rest frame $I$-band luminosity is approximately equivalent to the
observed $H$-band and so is calculated from the $J$ and $K$ photometry
(Table~1).  We co-add the $J,K$ images to obtain an aperture which is
used to extract the $J,K$ magnitudes and interpolate for the relevant
SED type.  We derive a rest-frame $I$-band magnitude of $M_{I} =
-24.1\pm 0.1$.  Correcting for the magnification due to the lens, the
absolute rest-frame $I$-band magnitude is M$_{I}^{rest} = -22.4 \pm
0.2$.  We apply the same technique for the rest-frame $B$-band
magnitude, (at $z\sim1$, $I$ and $R$-bands are the closest match to
rest-frame $B$-band) and estimate its uncertainty by computing the
magnitude for a variety of SED types which are consistent with the
observed optical colours.  We compute a corrected rest frame $B$-band
magnitude of $M_{B}^{rest} = -21.1 \pm 0.2$.  Accounting for the
$(1+z)^{4}$ dimming, the central surface brightness of the galaxy
($\mu_{b,_{0}}\sim20.6$) confirms $\#289$ as a high-surface brightness
(HSB) galaxy.

The inclination is determined by transforming the {\it HST} images of
the galaxy to the source plane.  We fit ellipses to an isophote of the
galaxy image in the source plane using the {\sc idl gauss2dfit}
routine. The ellipticity is then $e = 1 - b / a$ (where $a$ and $b$ are
the major and minor axis of the ellipse) and the inclination, $i$, is
$\cos i = b / a$.  The average axis ratio of the ellipses in the four
passbands (Table~1) is $0.43 \pm 0.07$ which translates to an
inclination of $64\pm 6$ degrees assuming an intrinsically circular
disk.

%
%
\begin{inlinefigure}
  \vspace*{0.5cm}
  \centerline{\psfig{file=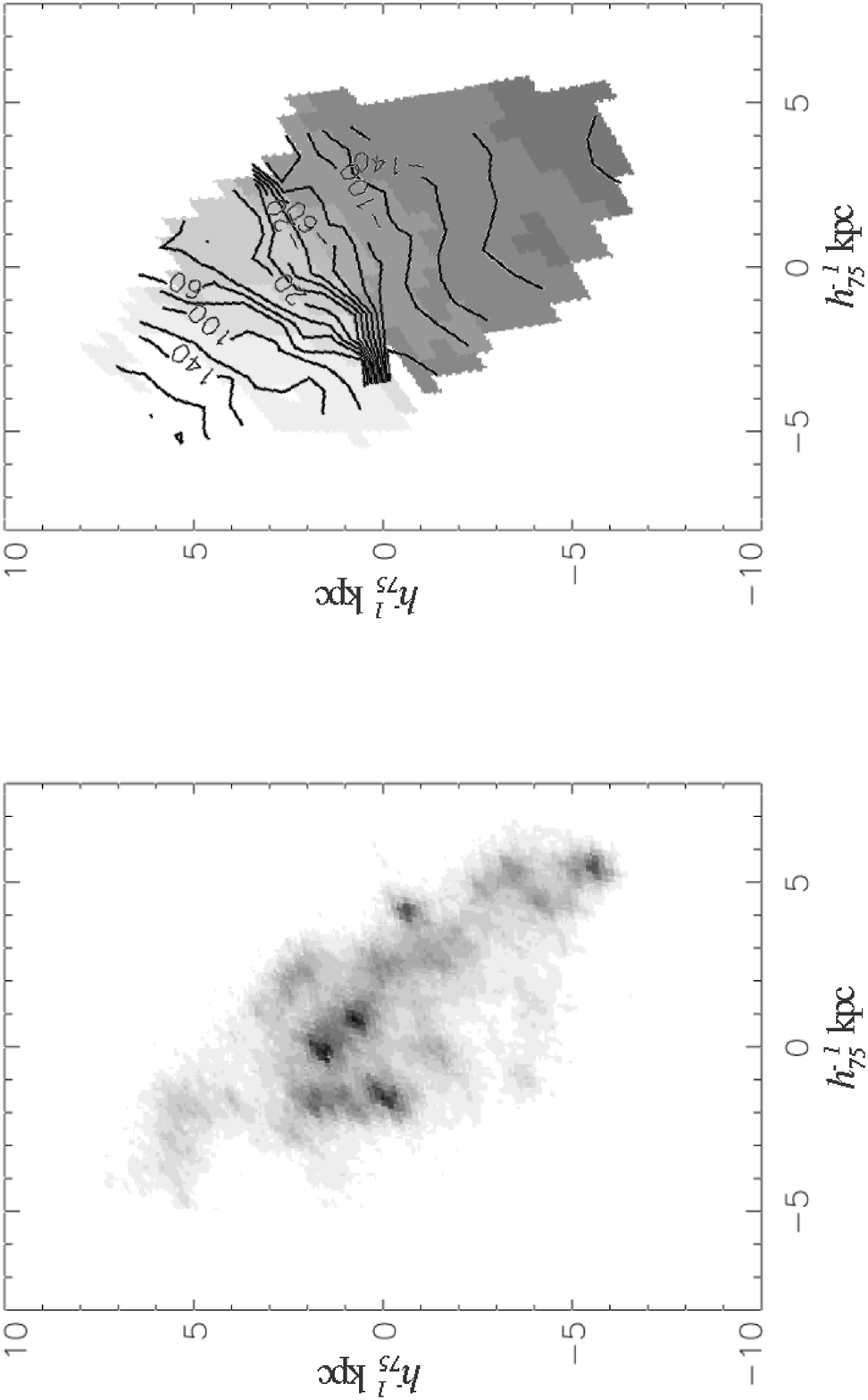,width=3.5in,angle=270}}
  \figcaption{ The reconstructed image of the arc corrected for lens
    magnification using the mass model of Smith et al.\ (2003).  Left:
    the reconstructed image of the galaxy based on {\it HST} imaging.
    Right: The velocity map of the galaxy in the source frame.  The
    dark and light regions represent redshift and blueshift
    respectively and the contours map the velocity in the source plane.
    The luminosity weighted magnification of the source is $4.92 \pm
    0.15$, but varies from $\sim5.6$ to $\sim4.9$ from the northern to
    southern end of the arc.  The scale shows the size of the galaxy in
    the source frame.  Without a lens, at $z=1$, $1''$ corresponds to
    $7.7$kpc.  \label{fig:source_image}}
\end{inlinefigure}


\vspace{0.5cm}
\subsection{GMOS Spectroscopic Imaging}

A\,2218-arc \# 289 was observed with the GMOS-IFU on Gemini North on
2002 June 12$^{\rm th}$ U.T.\ during Science Demonstration time for a
total of 5.4\,ks in $0.7''$ seeing and photometric conditions. The IFU
uses a lensed fiber system to reformat the $7'' \times 5''$ field into
two long slits (Allington-Smith et al.\ 2002).  Using an $I$-band
filter in conjunction with the R400 grating results in two tiers of
spectra recording a maximum field of view.  The spectral resolution of
this configuration is $\lambda/\Delta\lambda =2000$.  For the galaxy at
$z=1.034$, the emission for the [O{\sc ii}] doublet falls at a
wavelength of 7581$\AA$, in a region of low sky emission.

The GMOS data reduction pipeline was used to extract and wavelength
calibrate the spectra of each IFU element. The variations in
fiber-to-fiber response were removed in {\sc idl} by using continuum
regions either side of the expected range of [O{\sc ii}] emission.  The
[O{\sc ii}] doublet is clearly resolved in the GMOS spectra
(Fig.~\ref{fig:oii_spec}).  This is useful, since one of the [O{\sc
  ii}] lines can clearly be identified even when the other lies
immediately under the narrow sky line at $\sim7580\AA$.  The emission
line doublet was fitted using a $\chi^2$ minimisation procedure, taking
into account the greater noise at the position of the sky line.  The
spectra were averaged over a $3 \times 3$ spatial pixel region,
increasing this region to $4 \times 4$ pixels if the signal was too low
to give a sufficiently high $\chi^2$ improvement over a fit without the
line.  In regions where this averaging process still failed to give an
adequate $\chi^2$, no fit was made.  With a continuum fit we required a
minimum $\chi^2$ of 25 (S/N of 5) to detect the line.  To compute the
error, the velocity parameter of the best fit line is varied until the
signal drops by a $\chi^2$ of 9.  The range in velocity corresponds to
a formal 3$\sigma$ error.

We use the velocity field to infer the rotational velocity of the
galaxy's gas disk.  As can be seen from the major axis cross section,
(Fig.~\ref{fig:tf_plot} inset) the asymptotic rotation speed is
$v_{rot} = 186 \pm 16 \kms$, (i.e. the total velocity shift across the
galaxy is $2v_{rot}$).  The inclination corrected velocity of the
galaxy is 206 $\pm$ 17 $\kms$.  This corresponds to a dynamic mass of
$1.4\times 10^{11}M_{\odot} h_{75}^{-1}$ within a radius
$\sim14h^{-1}_{75}kpc$ in this cosmology.

In addition the shape of the rotation curve can be compared to the low
redshift ($z\sim0.03$) sample of spiral galaxies by Courteau (1997).
We fit the observed source plane 1-D rotation curve by convolving the
{\sc arctan} and multi-parameter models (Courteau 1997) with $0.7''$
seeing (transformed to the source plane).  Using a $\chi^{2}$ fit the
observed rotation curve is best described with an {\sc arctan} function
with transition radius, $r_{t}=1.7$kpc (where $r_{t}$ describes the
transition radius between the rising and flattening of the curve), or a
multi-parameter fit with $r_{t}=2.1$kpc, $\gamma=1.5$.  To compare the
shape of the rotation curve with Courteau's sample we compute
$r_{t}/r_{opt}$ (where $r_{opt}$ is the radius enclosing 83\% of the
light in the source plane photometry) to be $\sim0.25$.  This ratio is
higher than most galaxies in the local sample, but not anomalous.
Around 10\% of local bright galaxies have similarly shaped rotation
curves.

%
%
\begin{inlinefigure}
  \vspace*{0.5cm} \centerline{\psfig{file=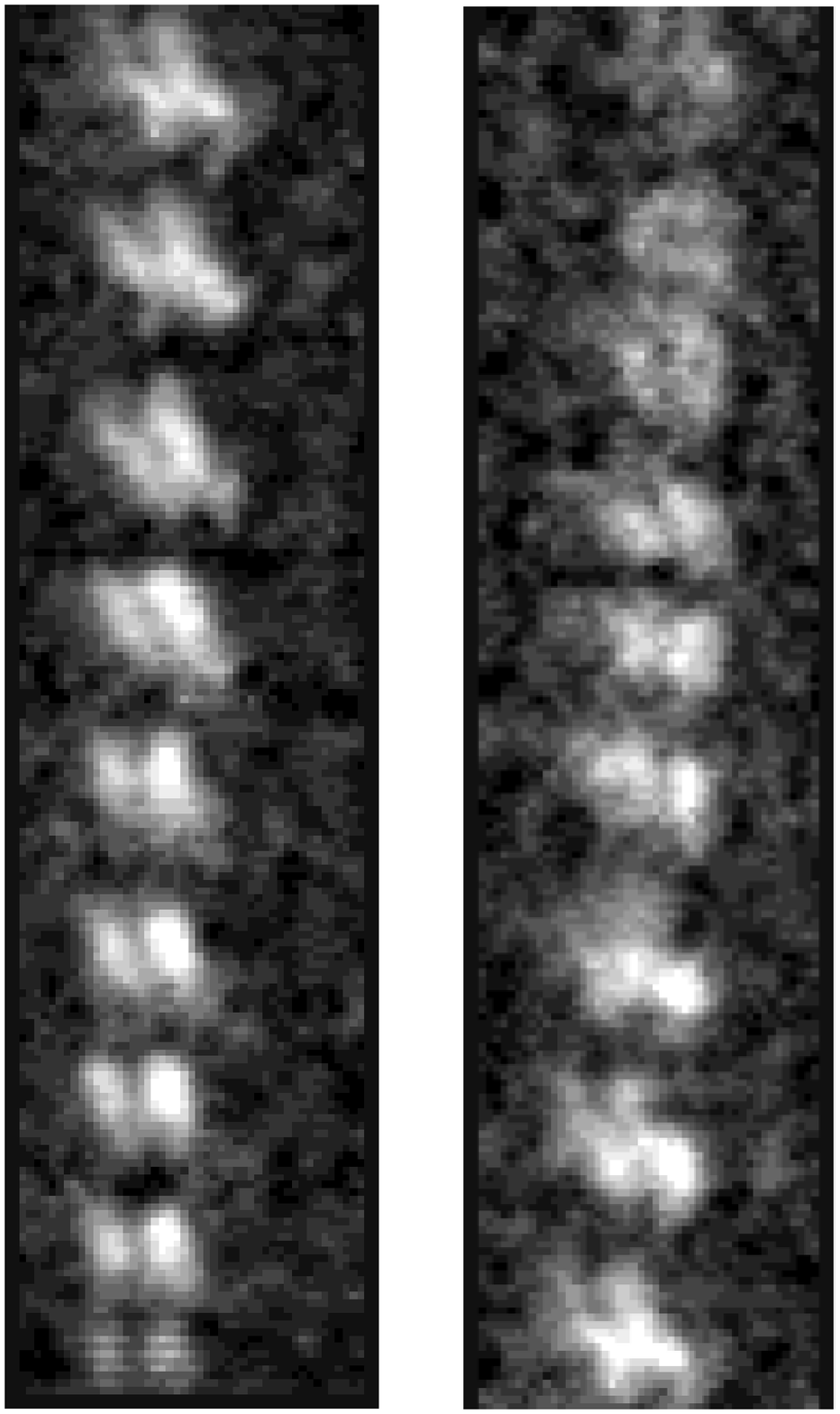,width=2.0in,angle=0}}
  \figcaption{ Examples of the [{O\sc ii}] emission doublet as seen in
    the reduced IFU data.  Each line of the image is the spectrum from
    a single IFU lenslet.  These are arranged in blocks, each block
    correspond to an East-West slice through the short dimension of the
    galaxy.  The rotation of the galaxy can be seen in each block as
    well as between blocks, clearly showing that the lensed galaxy's
    dynamics are resolved in both dimensions.
  \label{fig:oii_spec}}
\end{inlinefigure}


\section{Discussion}

%
%
\begin{figure*}[tbh]  
\centerline{\psfig{file=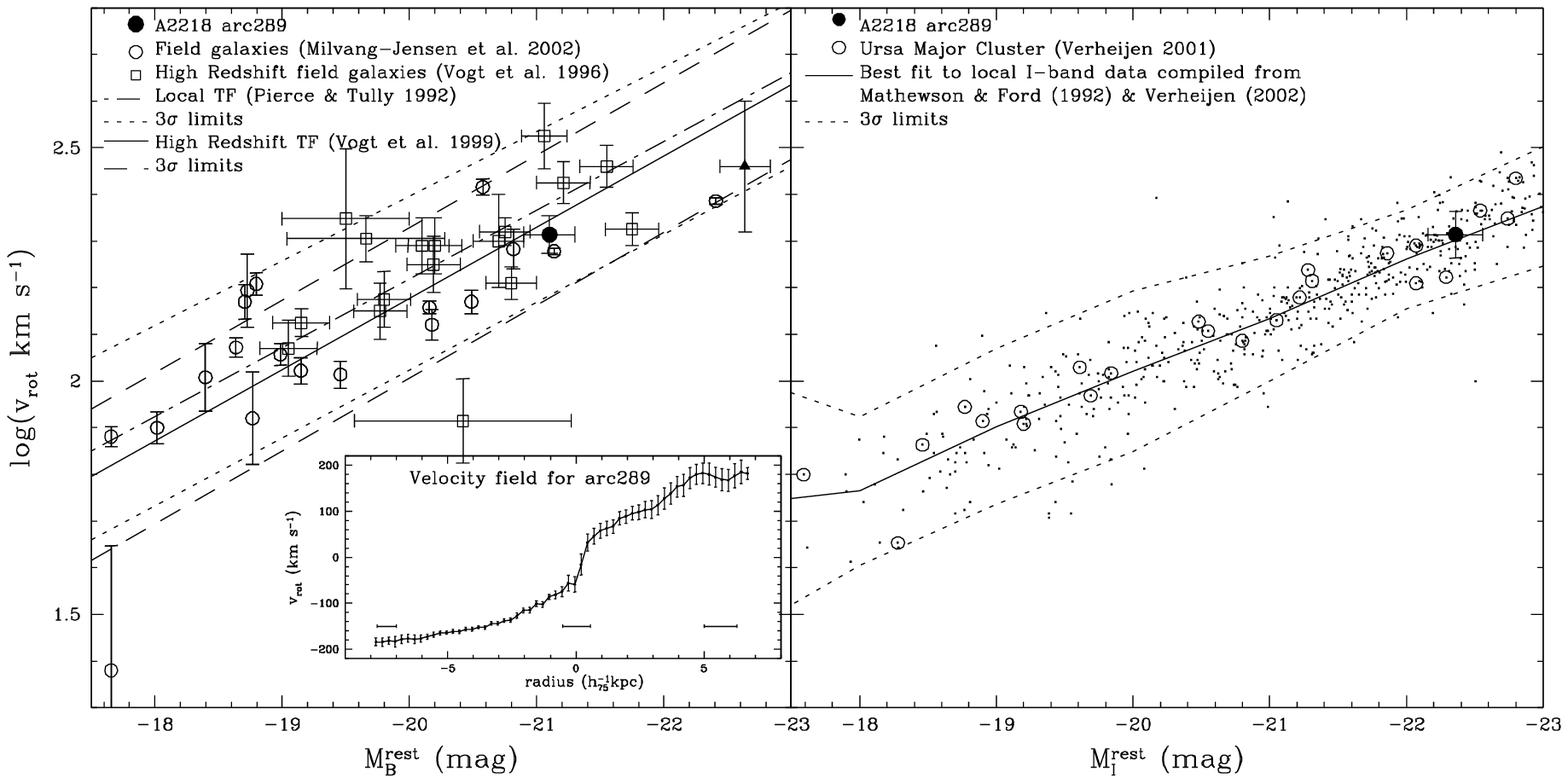,width=7.0in,angle=0}}  
\caption{\small Left: $Arc\#289$ on the Tully-Fisher relation in rest
  frame $B$-band compared to high redshift ($z\sim0.83$) field galaxies
  (Milvang-Jensen et al.\ 2001) and the high redshift sample from Vogt
  et al.\ (1999).  For comparison we show the low redshift local fit
  from Pierce \& Tully (1992).  The solid triangle shows a massive disk
  galaxy (L451) at $z=1.34$ (van Dokkum \& Stanford\ 2001).  The galaxy
  rotation curve (inset) shows the peak-to-peak rotation velocity of
  the arc in [O{\sc ii}] emission built from the IFU image in the
  source plane.  The error bars shown are formally 3-$\sigma$ and
  alternate points are independent. The horizontal error bars show
  $0.7''$ seeing transformed to the source plane.  Right: The rest
  frame $I$-band Tully-Fisher relation compiled from Mathewson \& Ford
  (1992) and from the Ursa-Major Cluster (Verheijen 2001). Arc\#289 is
  shown by the solid point and lies very close to the mean TF relation
  for present day spirals.  The small change in the $I$-band magnitude
  shown by arc\#289 suggests a preference for hierarchical rather than
  the ``classical'' formation model.}
\label{fig:tf_plot}
\end{figure*}


The tight correlation between luminosity and rotation velocity for
spiral galaxies in the local universe is known as the Tully-Fisher (TF)
relation (Tully \& Fisher\ 1977).  With a single high redshift galaxy,
we can measure the evolution of the offset in the TF relation under the
assumption that the slope remains fixed.  In Fig.~\ref{fig:tf_plot}, we
compare the rotation velocity and source brightness of arc \# 289 with
that of local and other high redshift galaxies in rest frame $B$ and
$I$.  Whilst the current high redshift data are concentrated in rest
frame $B$-band, rest frame $I$-band observations will prove a more
rigorous test of evolution of the TF relation since the corrections for
dust and on-going star formation are much smaller at longer
wavelengths. The rest frame $I$-band TF therefore gives a clearer
indication of the true stellar luminosity and hence the ratio of
stellar mass to total halo mass.

The position of the galaxy on the TF relation in both the $B$ and $I$
bands shows good agreement with local data (Pierce \& Tully 1992;
Mathewson \& Ford 1992; Haynes et al.\ 1999; Tully \& Pierce 2000;
Verheijen 2001). The galaxy is slightly offset to brighter magnitudes
in the $B$-band relation, as we would expect from its blue colors.
This data is in agreement with existing intermediate ($z\sim 0.5$, Vogt
et al.\ 1997; Milvang-Jensen et al.\ 2001; Bohm et al.\ 2002; Ziegler
et al.\ 2003) and high redshift studies ($z\sim~1$, Vogt et al.\ 2002;
Barden et al.\ 2003).  It should be noted, however, that these studies
generally have to infer the turn over in the rotation curve from
deconvolution of the rising rotation speed and the seeing.  This might
lead to a systematic underestimate of the asymptotic rotation speed.
In contrast, our data clearly resolve the shape of the rotation curve
(Fig.~\ref{fig:tf_plot} inset) leading to much smaller uncertainty in
the rotation speed of an individual galaxy.

Clearly, it would be dangerous to draw far reaching conclusions from a
single galaxy, but the following comparison illustrates the sensitivity
that can be expected if this level of evolution is confirmed by further
observations.  The theoretical evolution of the $I$-band TF relation
from hierarchical models of galaxy formation from Cole et al.\ (2000)
predict that for any given disk circular velocity, the $I$ band
luminosity should decrease by $\sim0.1$ magnitudes from $z=0$ to $z=1$,
whilst in the $B$-band such models predict an increase in luminosity of
$\sim0.5$ magnitude for the same redshift change.  In contrast, we
would expect a decrease in luminosity of $\sim0.7$ magnitudes from
$z=0$ to $z=1$ if all of the galaxy's mass were already in place at
$z=1$, but only half of the stars had yet formed. The small change in
the $I$-band magnitude shown by arc\#289 suggests a preference for
hierarchical rather than the ``classical'' formation model.

The methods we have developed in this paper show the potential of IFU
observations of gravitational lensed galaxies as a means of studying
the spatially resolved properties of high redshift galaxies in a
remarkable level of detail.  While we have concentrated here on the
dynamics of the galaxy \#289 in A\,2218, our data can also be used to
investigate the spatial distribution of star formation. By combining
these optical data with observations with near-infrared integral field
units, it will be possible to study the emission line ratios of [O{\sc
  ii}]:[O{\sc iii}] and H$\alpha$:H$\beta$.  We could then determine
the distribution of reddening across the galaxy and its spatially
resolved chemical abundance. This would provide powerful insight into
the nature of star forming galaxies at this early epoch.

\section{Acknowledgments}

Based on observations obtained at the Gemini Observatory, which is
operated by the Association of Universities for Research in Astronomy,
Inc., under a cooperative agreement with the NSF on behalf of the
Gemini partnership: the National Science Foundation (United States),
the Particle Physics and Astronomy Research Council (United Kingdom),
the National Research Council (Canada), CONICYT (Chile), the Australian
Research Council (Australia), CNPq (Brazil) and CONICET (Argentina).
We would like to thank Matt Mountain for accepting this programme for
GMOS IFU Science Demonstration; We thank Inger J{\o}rgensen, Matt
Mountain, Jean Rene-Roy, Kathy Roth, Marianne Takamiya, Roger Davies,
Gerry Gilmore, Bryan Miller, Chris Packman, Bo-Milvang-Jensen, Carlton
Baugh, David Gilank, Nathan Courtney and Richard McDermid for their
vital assistance with the planning, preperations, observations $\&$
discussion and Robert Content and Graham Murray for designing and
building the instrument.  AMS acknowledges the support of a PPARC
postgraduate studentship, RGB acknowledges support from the Leverhulme
Trust and Euro 3D Research Training Network, and IRS acknowledges
support from the Royal Society and the Leverhulme Trust.

\end{document}